\documentclass{article}
\usepackage{graphicx}
\usepackage{epsfig}
\usepackage{amsfonts}
\usepackage{amsmath, amssymb, array}
\usepackage{graphics,graphicx,float,mfpic}
\usepackage{epsfig}
\usepackage{color}
\newcommand{\be}{\begin{equation}}
\newcommand{\ee}{\end{equation}}
\newcommand{\bea}{\begin{eqnarray}}
\newcommand{\eea}{\end{eqnarray}}
\newcommand{\bmat}{\begin{pmatrix}}
\newcommand{\emat}{\end{pmatrix}}

\title{New stable phase of non uniform black strings in $\mbox{AdS}_d$}
\author{{\large T\'erence Delsate\footnote{terence.delsate@umh.ac.be} }\\ \\{\small Theoretical and Mathematical Physics Dept.}\\{\small University of Mons-Hainaut, Belgium}}
\date{\today}

\begin{document}
\maketitle

\begin{abstract}
We consider the non uniform $AdS$ black string equations in arbitrary number of dimension in a perturbative approach up to order $2$ and in a non perturbative. We restrict the study in the perturbative approach to the backreacting modes, since they provide the first relevant corrections on the thermodynamical quantities of the solutions. We also present some preliminary results in the construction of non-perturbative solutions, in particular, we present a first part of the non uniform - uniform black string phase diagram. Our results suggests the existence of a new stable phase for $AdS$ non uniform black strings, namely long non uniform black string, with the extra direction length of the order of the $AdS$ curvature. \\

\end{abstract}

\section{Introduction}

There has been a growing interest for solutions in AdS spacetime during the last decade, largely motivated by the AdS/CFT correspondence \cite{adscft2,adscft}. This correspondence relates the geometry of an asymptotically $AdS$ space with a conformal field theory defined on the background of the conformal boundary of the $AdS$ space. In this context, black objects are of great interest; for example, the Hawking-Page phase transition \cite{hptr} between the five dimensional spherically symmetric black holes and the thermal AdS background was interpreted by Witten, through AdS/CFT, as a thermal phase transition from a confining to a deconfining phase in the dual four dimensional $\mathcal N = 4$ super Yang-Mills theory \cite{whp}. 

In four dimensional asymptotically flat space, the uniqueness theorem guaranties that the horizon topology of black objects is $S_2$. In higher dimensions, this theorem does not hold and several black objects with various topology have already been constructed. Among them, let us mention the black strings \cite{solubs} of horizon topology $S_{d-3}\times S_1$, black rings \cite{solubr} (toroidal horizon topology), black holes ($S_{d-3}$ horizon topology) \cite{solubh} and their generalisation with charge, rotation, cosmological constant, etc.

In 1993, Gregory and Laflamme \cite{GL} showed that black strings are unstable toward long wavelength perturbations in asymptotically locally flat space  (see \cite{glrev} for a review). Just after that discovery, it was widely believed that the end point of the black string instability in asymptotically locally flat spacetime should be a caged black hole, where the asymptotical space in $d+1$ dimensions is $\mathcal M_d\times S_1$ (with $\mathcal M_d$ the $d$-dimensional Minkowski space), but it was argued \cite{Horowitz:2001cz} that the transition between a black string and a caged black hole would take an infinite proper time at the horizon and thus even 'more` for an observer lying outside the horizon. This motivated the study of extended black objects which are not translationally invariant along the $S_1$ coordinate, namely the non uniform black strings, first in a perturbative approach \cite{gubser} (see also \cite{kudoh}) and later in the full non linear regime \cite{Wiseman:2002zc}. The resulting thermodynamical phase diagram \cite{flatphase} is now well known although dynamical phase transitions are still to be constructed. In this diagram, the non uniform and the caged black hole branch meet at the merger point, where a topological phase transition is expected to occur.

 More recently, the question of the black strings instability in asymptotically locally $AdS$ spacetime has been addressed and it has been argued that this instability should persist in asymptotically locally AdS space in any number of dimensions for small AdS black strings \cite{rbd} where the ratio $r_h/\ell <<1$, with $r_h$ the horizon radius and $\ell$ the AdS radius, fixed by the cosmological constant $\Lambda$. Note that there exists another phase of AdS black strings, namely large AdS black strings \cite{rms,cophor}, where $r_h/\ell \geq 1$ which are thermodynamically and dynamically stable \cite{rbd}. As already noticed in \cite{hptr}, AdS space acts like a confining box: when the ratio $r_h/\ell$ becomes larger than some critical values, the wavelength of the instable modes cannot fit the 'AdS box' and thus the configuration is stable.

However, in asymptotically locally AdS spaces, much less is known about the counterpart of the solutions of $\Lambda = 0$ Kaluza-Klein black objects. The evolution of AdS black holes has been studied in \cite{gmbh} but still has to be done for AdS black strings. The thermodynamical properties of the uniform phase is now well known \cite{rms} and non uniform solution have been constructed in \cite{pnubsads} in a perturbative approach and in $6$ dimensions. In order to construct a phase diagram in AdS, one has to consider various stationary solutions and compare their thermodynamical properties since these stationary solutions should be the equilibrium configurations. This could give some clues on the endpoint of dynamical evolution and phase transitions between different black objects in AdS.

In this paper, we consider perturbative non uniform black strings in asymptotically locally $AdS$ spacetime for arbitrary number of dimensions. The decay of the non uniform modes in arbitrary number of dimension is sharp so it is difficult to construct the global conserved quantities. Instead, we focus on the horizon thermodynamical properties such as the entropy and the Hawking temperature in order to investigate the possible existence of new stable phases of non uniform black strings; namely long black strings, which should be counterparts of big $AdS$ black strings. It will turn out that this stable phase indeed exists and appears only in the non uniform phase. We also construct the non perturbative non uniform black string solution for some different values of the cosmological constant. 

This paper is organised as follows: in section 2, we present the model. In section 3, we give the near horizon and asymptotical solution to the equations while we present the boundary conditions and numerical technique in section 4. Section 5 is devoted to the thermodynamical properties of the solutions, we present the numerical results and give a discussion of the solutions in section 6. Finally we present the solution to the non perturbative problem in section 7 and discuss the properties of the non uniform black string in the regime where the cosmological constant is small. The results are summarized in the conclusion.

\section{The model}
We consider the $d$-dimensional Einstein-Hilbert action supplemented with the Hawking-Gibbons boundary terms and a negative cosmological constant:
\be
S = \frac{1}{16\pi G} \int_{\mathcal{M}} \sqrt{-g}\left(R + \frac{(d-1)(d-2)}{\ell^2}  \right)d^dx + \frac{1}{8\pi G}\int_{\partial\mathcal{M}}\sqrt{-h} Kd^{d-1}x,
\label{seh}
\ee
where $G$ is the $d$-dimensional Newton constant, $g$ is the determinant of the metric on the spacetme manifold $\mathcal M$, $h$ the determinant of the induced metric on the boundary manifold $\partial\mathcal M$, $R$ is the scalar curvature, $\ell$ is the $AdS$ radius and $K$ is the trace of the extrinsic curvature of the boundary spacetime. In what follows, we will work in geometrical units where $G=1$.

We supplement the action \eqref{seh} with the following ansatz, which is relevant in the study of non uniform black strings \cite{gubser}:
\be
ds^2 = -b(r)e^{2A(r,z)}dt^2 + e^{2B(r,z)}\left(\frac{dr^2}{f(r)} + a(r)dz^2\right) + e^{2C(r,z)}r^2d\Omega_{d-3}^2,
\label{ansnubs}
\ee
where $d\Omega_{d-3}^2$ is the square of the line element on the unit $(d-3)$-sphere and $z\in[0,L]$, $L$ being a real number.

The ansatz \eqref{ansnubs} leads to partial differential equations which are technically difficult to deal with. In order to obtain a set of ordinary differential equations, we develop the non uniformity (the $z$ depending part) in a Fourier series and in term of a small parameter $\epsilon$:
\be
X(r,z)= \epsilon X_1(r)\cos(kz) +\epsilon^2\left( X_0(r) + X_2(r)\cos(2kz) \right) + \mathcal O(\epsilon)^3,
\label{decomp}
\ee
where $X$ generically denotes $A,B,C$ and $k=2\pi/L$.

With the expansion \eqref{decomp}, we obtain ordinary differential equations at each order in $\epsilon$ and for each independent Fourier mode. The equations at order $\epsilon^0$ are the background equations; the solutions of these equations have been studied in \cite{rms}. The order $\epsilon^1$ leads to the linear stability of the background and the solutions have been constructed in \cite{rbd}.

In order $\epsilon^2$, there are two independent modes: the $X_0$ and the $X_2$ fields. The $X_0$ modes are the backreacting fields and have been studied in \cite{pnubsads} for $d=6$. They are relevant for the first corrections on thermodynamical quantities of non uniform black strings. The $X_2$ modes are massive modes and do not contribute in thermodynamical corrections (at first relevant corrections)\footnote{thermodynamical quantities involve an integration over $z$ from $0$ to $2\pi/L$, thus suppressing $X_2$ terms, see \cite{pnubsads}.}. In addition, the $X_2$ modes are independent of the $X_0$ fields since they are different Fourier modes; that's the reason why we will consider only the $X_0$ fields at order $\epsilon^2$ in this paper.

The equations for the background can be found in \cite{rms} and the equations for the first order are in \cite{rbd}. The equations for the second order in $\epsilon$ are quite long and we prefer not writing them here but they are straightforward to obtain. Let us just remind that it is possible to solve algebraically for the first order field $B_1$ (resp. second order field $B_0$ ) as a function of the other first order field (resp. second order fields). Let mention also that the equations for the backreacting fields present the same shift invariance than in the $6$-dimensional case, $X_0\rightarrow X_0+\mbox{const}$.

\section{Near-horizon and asymptotic expansion}

In order to compute the asymptotic expansion, two cases must be distinguished: the case of odd number of dimensions and the case of even number of dimensions. In odd number of dimension, $log$ terms arise in the background asymptotic expansion while in even number of dimensions, there are no $log$ terms in the expansion; these terms might have a nontrivial repercussion on the first and second order fields, but it turns out that \emph{at the leading order} of the asymptotic expansion, it is not the case.

The background fields obey the Fefferman-Graham expansion which can be found in \cite{rms} for the case of interest. Let us just give the leading order expansion of the background fields:
\be
a(r) \approx b(r) \approx f(r) = \frac{r^2}{\ell^2} + \mathcal O (1),
\label{bgas}
\ee
which is the appropriate asymptotic behaviour of a locally asymptotically $AdS$ spacetime.

 In such an asymptotic background, the leading order of the asymptotic expansion for the first order fields is given by:
\bea
A_1(r) &=& -(d-3)\gamma_1 \left(\frac{\ell}{r}\right)^{d-1}+ \mathcal O\left(\frac{\ell}{r}\right)^{d+2} \ ,\ C_1(r) = \gamma_1 \left(\frac{\ell}{r}\right)^{d-1} + \mathcal O\left(\frac{\ell}{r}\right)^{d+2},
\eea
with $\gamma_1$ a real constant to be determined numerically. The influence of the log terms we mentioned shows up in higher order terms in the expansion and is not relevant for our analysis. In fact, this is not the most general asymptotic expansion for the first order fields. The most general expansion contains constant terms and terms of order $r^{-(d-4)}$. These terms cancel once the boundary condition is imposed ($A_1(r)\rightarrow0,\ C_1(r)\rightarrow0$ when $r\rightarrow\infty$; see next section).

The backreacting fields $A_0,C_0$ follow the same pattern:
\bea
A_0(r) &=& -(d-3)\gamma_0 \left(\frac{\ell}{r}\right)^{d-1}+ \mathcal O\left(\frac{\ell}{r}\right)^{d+2} \ ,\ C_0(r) = \gamma_0 \left(\frac{\ell}{r}\right)^{d-1} + \mathcal O\left(\frac{\ell}{r}\right)^{d+2},
\label{bras}
\eea
where $\gamma_0$ is a real constant, also to be determined numerically. Once again, in the most general expansion, there are lower order terms which would give infinite contribution to the mass and thus are unphysical \cite{pnubsads}. 

Higher order terms for first and second order fields can be obtained after a straightforward calculation.

The near horizon behaviour of the background fields and first order are reminded here:
\bea
a(r) &=& a_h + \frac{2a_h\left(d-1\right)r_h}{\left(d-4\right)\ell^2+\left(d-1\right)r_h^2}(r-r_h) + \frac{2a_h\left(d-1\right)^2r_h^2}{\left(\left(d-4\right)\ell^2+\left(d-1\right)r_h^2 \right)^2}\frac{(r-r_h)^2}{2}+\mathcal O(r-r_h)^3,\nonumber\\ 
b(r) &=& b_h(r-r_h) + \frac{b_h\left(d-4\right)\left(\left(d-3\right)\ell^2+\left(d-1\right)r_h^2 \right)}{\left(d-4\right)\ell^2r_h + \left(d-1\right)r_h^3}\frac{(r-r_h)^2}{2}+\mathcal O(r-r_h)^3,\\ 
f(r) &=& \left(\frac{(d-1))}{r_h} + \frac{(d-4)r_h}{\ell^2}\right)(r-r_h) - (d-4)\left(\frac{d-1}{\ell^2}+\frac{d-3}{r_h^2}\right)\frac{(r-r_h)^2}{2}+ \mathcal O(r-r_h)^3;\nonumber\\
A_1(r) &=& A_{10} + A_{11}(r-r_h) + \mathcal O (r-r_h)^2\ ,\ C_1(r) = C_{10} + C_{11}(r-r_h) + \mathcal O (r-r_h)^2,
\eea
where
\bea
A_{11} &=& -\frac{2a_h\left(A_{10}-C_{10}\right)\left(d-4\right)\left(d-3\right)\ell^2+ \left( 2a_hA_{10}\left(d-5\right)\left(d-1\right)+
\left(-2A_{10}+C_{10}\left(d-3\right)  \right)k^2\ell^2\right)r_h^2}{3a_hr_h\left(\left(d-4\right)\ell^2+\left(d-1\right)r_h^2 \right) }\nonumber\\
C_{11} &=& \frac{2\left(A_{10}-C_{10}\right)}{r_h} + 
  \frac{C_{10}\left(2a_h\left(d-1\right)+k^2\ell^2\right)r_h}{a_h\left(d-4\right)\ell^2+a_h\left(d-1\right)r_h^2},
\eea
and $a_h, b_h$ are normalisation constants to be fixed such that the background fields follow \eqref{bgas}; $A_{10}, C_{10}$ are real constants. Note that due to the linearity of the first order equation, either $A_{10}$ or $C_{10}$ can be fixed arbitrarily, only the ratio $A_{10}/C_{10}$ is non arbitrary.

The second order fields have the following near-horizon expansion:
\bea
A_0(r)= A_{00} + A_{01}(r-r_h) + \mathcal O(r-r_h)^2\ ,\ C_0(r) = C_{00} + C_{01}(r-r_h) +  \mathcal O(r-r_h)^2,
\eea
where $A_{00},C_{00}$ are to be fixed using the invariance under $A_0\rightarrow A_0+\mbox{const},\ C_0\rightarrow C_0+\mbox{const}$ such that the fields $A_0,C_0$ decay to $0$ at infinity, $C_{01}$ is an arbitrary real constant and 
\be
A_{01} = -\frac{a_hC_{01}\left(d-4\right)\left(d-3\right)\ell^2 + \left(A_{10}^2+ 2A_{10}C_{10}\left(d-3\right)- C_{10}^2\left(d-3\right)\right)k^2\ell^2r_h+a_hC_{01}\left(d-3\right)\left(d-1\right)r_h^2}{3a_h\left(d-4\right)\ell^2+ 
      3a_h\left(d-1\right)r_h^2}.
\ee

This expression for $A_{01}$ is obtained by imposing regularity of the equation in the near horizon region; if $A_{01}$ is not chosen to be the above expression, there will exist a diverging term proportional to $(r-r_h)^{-1}$ in the near horizon limit.

\section{Boundary conditions and numerical technique}

The numerical technique follows that of reference \cite{pnubsads}: we first integrate the background fields with appropriate boundary conditions (see ref. \cite{rms}) then we integrate the first order fields (see \cite{rbd} for the boundary conditions). The backreacting fields are then integrated with the following initial conditions:
\bea
A_0(r_h) &=& \alpha_0,\ \ C(r_h) = \chi_0,\ \ C'(r_h)=\chi_1\\
A_0'(r_h)&=&-\frac{a_h\chi_1\left(d-4\right)\left(d-3\right)\ell^2 + \left(A_{10}^2+ 2A_{10}C_{10}\left(d-3\right)- C_{10}^2\left(d-3\right)\right)k^2\ell^2r_h+a_h\chi_1\left(d-3\right)\left(d-1\right)r_h^2}{3a_h\left(d-4\right)\ell^2+ 
      3a_h\left(d-1\right)r_h^2},\nonumber
\eea
where $\alpha_0,\chi_0$ are arbitrary constant to be fixed a posteriori using the shift invariance $A_0(r)\rightarrow A_0(r)+\mbox{const}$, $C_0(r)\rightarrow C_0(r)+\mbox{const}$, while $\chi_1$ is a constant which is tuned such that the fields $A_0,C_0$ follow the decay \eqref{bras}.

In practice, we integrate the background and the first order with the solver \emph{Colsys} \cite{COLSYS} and the second order is integrated using a Runge Kutta algorithm at order 4. The integration is carried out from the horizon radius, $r_h$ to some $R>>r_h$. The fields $A_0,C_0$ follow the asymptotic \eqref{bras} if $R C_0'(R) + (d-1) C_0(R) = 0$. The problem here is that $C_0$ can always be shifted by an arbitrary constant so we impose the decay \eqref{bras} transposed to the derivative of $C_0$:
\be
RC_0''(R) + dC_0'(R) = 0,
\label{brbc}
\ee
the value of $C_0''(R)$ being obtained using the field equations.

The value of $RC_0''(R) + dC_0'(R)$ is a function of $\chi_1$, say $\mathcal C(\chi_1)$. The backreacting fields will follow the asymptotic decay \eqref{bras} if $\chi_1$ is chosen to be a root of $\mathcal C(\chi_1)$. We used the Newton algorithm in order to find the value of $\chi_1$ such that $\mathcal C(\chi_1) = 0$.


\section{Thermodynamical properties}
The entropy is given by a quarter of the event horizon area and the first relevant correction to the entropy coming from the non uniformity appears at the backreacting level. Let us write the entropy as $S = S_0 + \epsilon^2\delta S$, $S_0$ being the entropy of the uniform black string, $\delta S$ the correction from the backreaction. Then, 
\bea
S_0 &=& \frac{2\pi}{k}\frac{\Omega_{d-3}r_h^{d-3} \sqrt{a_h}}{4} \\
\delta S/S_0 &=& \frac{a_h\left(-A_{10}^2+2A_{10}C_{10}(d-3)+\left( 4C_{00}+C_{10}^2(d-3)  \right)(d-3)  \right)(d-1)+C_{10}(A_{10}+C_{10})(d-3)k^2\ell^2}{4a_h(d-1) },\nonumber
\eea
where $\Omega_{d-3}$ is the surface of the unit $(d-3)$-sphere and the background quantities are evaluated with the length $L = 2\pi/k$, $k$ being determined from the solution to the first order equations.

The temperature at the horizon can be computed demanding regularity in the Euclidean sections and is given by:
\bea
T_H &=& \frac{1}{4\pi}\sqrt{\frac{b_1}{r_h \ell^2}\left((d-1)r_h^2+(d-4)\ell^2\right)},\\
\delta T_h/T_H &=&\frac{2a_0(d-1)\left( 2A_{00} + A_{10}^2\right) + (d-3)(A_{10}-C_{10})C_{10}k^2\ell^2}{4a_0(d-1)},\nonumber
\eea
where $T_H$ is the background temperature and $\delta T_H$ the corrections from the backreaction, the temperature of the non uniform phase being denoted by $T_H + \epsilon^2\delta T_H$.

The mass and tension can be computed using the counterterm procedure \cite{counter}. We refer the reader to \cite{rms,pnubsads} for the details of the technique and an application to uniform asymptotically $AdS$ black string and perturbative non uniform $AdS$ black strings in $6$ dimensions. Let us notice that there exists another procedure, free of ambiguities that might arise during the standard counterterm procedure, namely the Kounterterm procedure \cite{kounter}. 

However, the thermodynamical quantities involved in these procedures are difficult to extract from the numerical solution in the general $d$-dimensional case because of the backreacting fields asymptotical decay \eqref{bras}. Instead, it is in principle possible to compute the mass by integrating the first law with fixed asymptotical length ($L$ fixed) and to use the Smarr formula \cite{rms} to extract the tension. In practice, this procedure works well except for very small value of the horizon radius (with $AdS$ length fixed to one), which is the region of interest in the perturbative approach. We shall come back to this point in the discussion. For our purpose, it is sufficient to evaluate only the entropy and Hawking temperature (with corrections arising form the backreactions).

In order to investigate the thermodynamical stability of the non uniform phase, we consider the specific heat, defined as the derivative of the entropy with respect to the temperature. In the case of perturbative non uniform black strings at order $\epsilon^2$, the specific heat reduces to
\be
C_p^{NU} = \frac{\delta S}{\delta T_H},
\ee
with $\delta S$ and $\delta T_H$ as defined above.

This quantity gives the variation of entropy whith respect to the Hawking temperature in the non uniform phase at the merger point; non uniform solutions with $C_p<0$ are thermodynamically unstable, while solutions with $C_p >0$ are thermodynamically stable. Consequently, in order to investigate the existence of a stable phase, it is sufficient to compare the sign of $\delta S$ and $\delta T_H$.

\section{Numerical results and discussion}

\begin{figure}[H]
\begin{center}
\includegraphics[scale=.7]{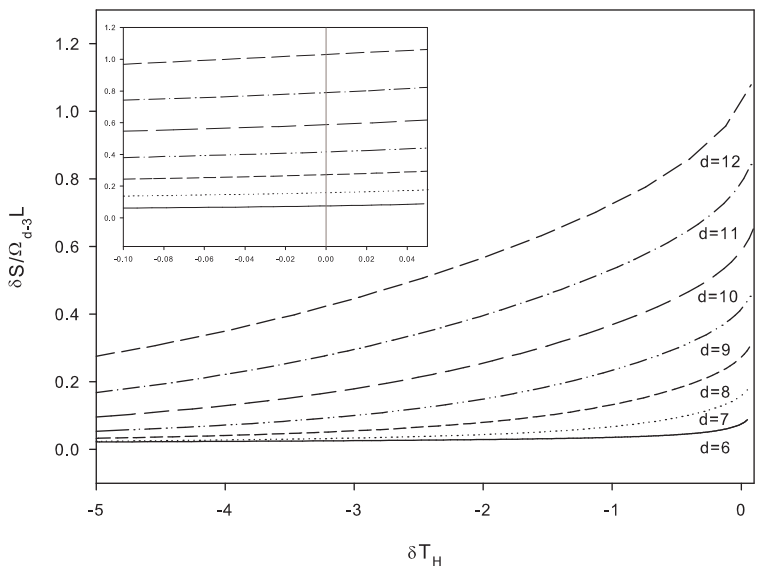}	
\caption{The correction on the entropy $\delta S$ as a function of the correction on the Hawking temperature $\delta T_H$ for various number of dimensions. There are points on this figure with $\delta T_H >0$ corresponding to large value of $\mu_1$. This stable phase occurs for small value of $k$ (i.e. large value of $L$) and is the analogue of the big $AdS$ black string, with the large length being now in the $S^1$ direction. The small figure in upper left box is a zoom of the region with $\delta T_H>0$.}
\label{dSdT}
\end{center}
\end{figure}

We integrated the field equations for values of $d$ from $5$ to $15$. We considered only perturbative non uniform black strings with critical length. The critical length is well defined only for small $AdS$ black strings (for big $AdS$ black strings, the critical wavenumer $k$ becomes imaginary and so does the critical length $L=2\pi/k$). Our results show that for every number of dimensions considered, there is a phase of non uniform black string presenting a positive specific heat (Fig. \ref{dSdT}).

This phase arises for small values of the critical wave number, i.e. large value of the length (with $\ell$ fixed). This is the analogue of the small-big black strings or black hole in $AdS$ where the relevant parameter for the thermodynamical stability was $r_0/\ell$. Here, the relevant parameter is $\mu_1 = L/\ell$; short non uniform black strings solutions with $\mu_1<<1$ are thermodynamically unstable while long non uniform black strings with $\mu_1\approx1$ are thermodynamically stable.

\begin{table}[h]
	\centering
			$\begin{array}{ccccccccc}
			\hline
			\hline
			  d	     &\vline &	6	      &	7	      &			 8  &	9	      &	10	      &	11	      &	12      \\
			  \hline
	     \mu_1^{cs}&	\vline &	8.66409	&	8.76464	&	9.02511	&	9.3052	&	9.56981		&	9.80617		&  10.1825\\
	  \hline
			\hline
	\end{array}$
	\caption{The value of the ratio $\mu_1^{cs}=L/\ell$ where the stable phase occurs for various $d$.}
	\label{tab:muc}
\end{table}

Note that this effect is not present for the uniform phase. In the uniform phase, the length enters only as an overall factor in the thermodynamical quantities. In the non uniform phase, the thermodynamical quantities depend on the length in a non trivial way (via $k=2\pi/L$). 
Let us stress the fact that these short-long black strings are present in the small $AdS$ phase, since it is the phase we are dealing with.
We expect this property to be a generic feature of non uniform black strings in arbitrary number of dimensions.\\

\begin{figure}[H]
\begin{center}
\includegraphics[scale=.7]{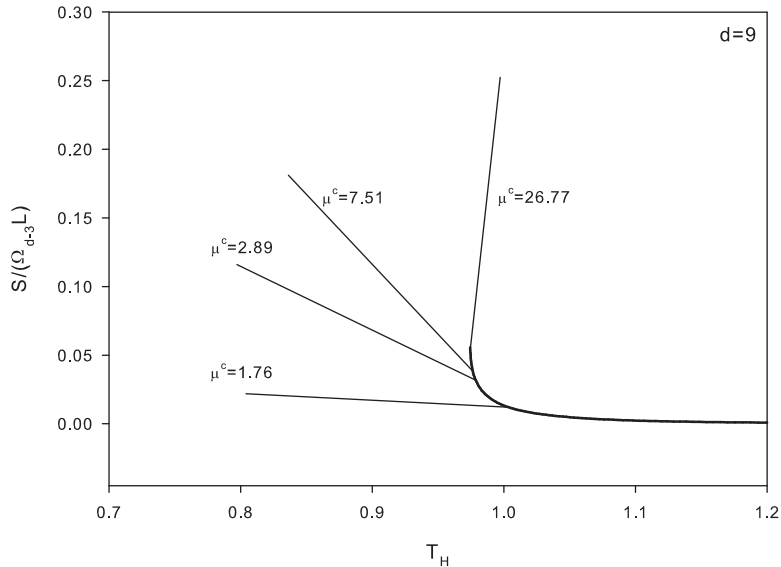}
\caption{The phase diagram in the $S-T_H$ plane for $d=9$. The background and the direction of the corrections for some value of the rescaled length are shown. The same pattern holds for all the number of dimension we considered.}
\label{phase}
\end{center}
\end{figure}

Figure \ref{phase} shows the direction of the phase transition in a $S-T_H$ diagram for $d=9$. In the short black string phase (i.e. with $L<<\ell$), the entropy increases and the temperature decreases along the non uniform phase while in the long black string phase ($L\approx\ell$) the entropy increases and so does the temperature. Let us emphasize once again that this new phase of non uniform black strings occurs for small black strings ($r_0<< \ell$) since the length is not well defined for big black strings ($r_0\approx \ell$).

We wanted to compute the mass and tension for the solutions we constructed in order to investigate the existence of a ($\ell$-depending\footnote{with fixed horizon radius or equivalently $r_0$ depending, with fixed $\ell$.}) critical dimension, but it turned out that these quantities are much difficult to evaluate precisely. The problem comes from the small values of the horizon radius: when integrating the background, we impose generically $a(r_h)=1,\ b(r_h)=f(r_h)=0$, and use the arbitrariness of the $a,b$ functions normalisation in order to impose $a,b,f\rightarrow r^2/\ell^2$ asymptotically, a posteriori. For small value of $r_h$, the asymptotical region appears for larger and larger value of the radial coordinate; this introduces numerical noise in the determination of the normalisation factors for the background functions and thus affects the horizon quantities. Moreover, the small black string phase is defined for $r_h<1$ (we are working with $\ell=1$; fixing another value of $\ell$ doesn't solve the problem.). The investigation of a critical dimension thus deserves further attention. 

However, we expect this long non uniform black string phase to be relevant in the potential existence of a $\ell$-depending critical dimension, since there is a major change in the thermodynamical properties of this phase.

\section{Non Uniform Black String}
In this section we present some preliminary results on the solution to the full non-linear system of partial differential equations in six dimensions. These equations can obtained starting from the ansatz \eqref{ansnubs}, without developing the non uniformity as a Fourier Series. However, demanding regularity on the horizon leads to non trivial relations between the functions $A,B,C$. 

Changing the parametrisation of the background (where $A=B=C=0$) to a conformal-like gauge, 
\be
ds^2 = -\tilde b(\tilde r)dt^2 + \frac{\tilde r^2d\tilde r^2}{(r_h^2+\tilde r^2)\tilde f(\tilde r)} + \tilde a(\tilde r)dz^2 + (r_h^2 + \tilde r^2)d\Omega_{3}^2,
\ee
and using a non uniform ansatz in this gauge,
\be
ds^2 = -\tilde b(\tilde r)e^{2\tilde A(\tilde r,z)}dt^2 + e^{2\tilde B(\tilde r,z)}\left(\frac{\tilde r^2d\tilde r^2}{(r_h^2+\tilde r^2)\tilde f(\tilde r)} + \tilde a(\tilde r)dz^2\right) + (r_h^2 + \tilde r^2)e^{2\tilde C(\tilde r,z)}d\Omega_{d-3}^2,
\label{ansnubs2}
\ee
leads to much simpler regularity conditions at the horizon, which will be given later.

In this parametrisation, the horizon is located at $\tilde r=0$ and the functions $\tilde b$ and $\tilde f$ behave like $\tilde r^2$ close to the horizon. Note that the variable $\tilde r$ and the functions $\tilde a,\tilde b,\tilde f$ are related to the variable $r$ and the functions $a,b,f$ according to
\be
r = \sqrt{r_h^2 + \tilde r^2}\ ,\ \tilde a(\tilde r)=a(r)\ ,\ \tilde b(\tilde r)=b(r)\ ,\ \tilde f(\tilde r)=f(r).
\ee

The equations with the parametrisation \eqref{ansnubs2} are given by

\begin{eqnarray}
&& \frac{-5e^{2\tilde B(\tilde r,z)}r^2}{s^2\tilde f(\tilde r)g(\tilde r)} - \frac{\tilde b'(\tilde r)}{2r\tilde b(\tilde r)} + 
  \frac{\tilde a'(\tilde r)\tilde b'(\tilde r)}{4\tilde a(\tilde r)\tilde b(\tilde r)} - \frac{{\tilde b'(\tilde r)}^2}{4{\tilde b(\tilde r)}^2} + 
  \frac{\tilde b'(\tilde r)\tilde f'(\tilde r)}{4\tilde b(\tilde r)\tilde f(\tilde r)} + \frac{\tilde b'(\tilde r)g'(\tilde r)}{\tilde b(\tilde r)g(\tilde r)} \nonumber \\
&+& \frac{\tilde b''(\tilde r)}{2\tilde b(\tilde r)} + 
  \frac{r^2{\tilde A^{(0,1)}(\tilde r,z)}^2}{L^2\tilde a(\tilde r)\tilde f(\tilde r)g(\tilde r)} + 
  \frac{3r^2\tilde A^{(0,1)}(\tilde r,z)\tilde C^{(0,1)}(\tilde r,z)}{L^2\tilde a(\tilde r)\tilde f(\tilde r)g(\tilde r)} + 
  \frac{r^2\tilde A^{(0,2)}(\tilde r,z)}{L^2\tilde a(\tilde r)\tilde f(\tilde r)g(\tilde r)} - \frac{\tilde A^{(1,0)}(\tilde r,z)}{r} \nonumber \\
&+&  \frac{\tilde a'(\tilde r)\tilde A^{(1,0)}(\tilde r,z)}{2\tilde a(\tilde r)} + \frac{\tilde b'(\tilde r)\tilde A^{(1,0)}(\tilde r,z)}{\tilde b(\tilde r)} +
  \frac{\tilde f'(\tilde r)\tilde A^{(1,0)}(\tilde r,z)}{2\tilde f(\tilde r)} + \frac{2g'(\tilde r)\tilde A^{(1,0)}(\tilde r,z)}{g(\tilde r)} \nonumber \\
&+&  {\tilde A^{(1,0)}(\tilde r,z)}^2 + \frac{3\tilde b'(\tilde r)\tilde C^{(1,0)}(\tilde r,z)}{2\tilde b(\tilde r)} + 
  3\tilde A^{(1,0)}(\tilde r,z)\tilde C^{(1,0)}(\tilde r,z) + \tilde A^{(2,0)}(\tilde r,z)=0,
\label{eqnu1}
\end{eqnarray} 
\begin{eqnarray}
&& \frac{3e^{2\tilde B(\tilde r,z) - 2\tilde C(\tilde r,z)}r^2}{\tilde f(\tilde r){g(\tilde r)}^2} + 
  \frac{5e^{2\tilde B(\tilde r,z)}r^2}{s^2\tilde f(\tilde r)g(\tilde r)} - \frac{\tilde a'(\tilde r)}{2r\tilde a(\tilde r)} - 
  \frac{{\tilde a'(\tilde r)}^2}{4{\tilde a(\tilde r)}^2} + \frac{\tilde a'(\tilde r)\tilde f'(\tilde r)}{4\tilde a(\tilde r)\tilde f(\tilde r)}  \nonumber \\
&+&  \frac{\tilde a'(\tilde r)g'(\tilde r)}{4\tilde a(\tilde r)g(\tilde r)} - \frac{3\tilde b'(\tilde r)g'(\tilde r)}{4\tilde b(\tilde r)g(\tilde r)} - 
  \frac{3{g'(\tilde r)}^2}{4{g(\tilde r)}^2} + \frac{\tilde a''(\tilde r)}{2\tilde a(\tilde r)} - 
  \frac{3r^2\tilde A^{(0,1)}(\tilde r,z)\tilde C^{(0,1)}(\tilde r,z)}{L^2\tilde a(\tilde r)\tilde f(\tilde r)g(\tilde r)} - 
  \frac{3r^2{\tilde C^{(0,1)}(\tilde r,z)}^2}{L^2\tilde a(\tilde r)\tilde f(\tilde r)g(\tilde r)} \nonumber \\
  &+& 
  \frac{r^2\tilde B^{(0,2)}(\tilde r,z)}{L^2\tilde a(\tilde r)\tilde f(\tilde r)g(\tilde r)} -
\frac{3g'(\tilde r)\tilde A^{(1,0)}(\tilde r,z)}{2g(\tilde r)} - 
  \frac{\tilde B^{(1,0)}(\tilde r,z)}{r} + \frac{\tilde a'(\tilde r)\tilde B^{(1,0)}(\tilde r,z)}{2\tilde a(\tilde r)}  \nonumber \\
&+&  \frac{\tilde f'(\tilde r)\tilde B^{(1,0)}(\tilde r,z)}{2\tilde f(\tilde r)} + \frac{g'(\tilde r)\tilde B^{(1,0)}(\tilde r,z)}{2g(\tilde r)}
- 
  \frac{3\tilde b'(\tilde r)\tilde C^{(1,0)}(\tilde r,z)}{2\tilde b(\tilde r)} - \frac{3g'(\tilde r)\tilde C^{(1,0)}(\tilde r,z)}{g(\tilde r)}  \nonumber \\
&-& 
  3\tilde A^{(1,0)}(\tilde r,z)\tilde C^{(1,0)}(\tilde r,z) - 3{\tilde C^{(1,0)}(\tilde r,z)}^2 + \tilde B^{(2,0)}(\tilde r,z)=0,
\label{eqnu2}
\end{eqnarray} 
\begin{eqnarray}
&&\frac{-2e^{2\tilde B(\tilde r,z) - 2\tilde C(\tilde r,z)}r^2}{\tilde f(\tilde r){g(\tilde r)}^2} - 
  \frac{5e^{2\tilde B(\tilde r,z)}r^2}{s^2\tilde f(\tilde r)g(\tilde r)} - \frac{g'(\tilde r)}{2rg(\tilde r)} + 
  \frac{\tilde a'(\tilde r)g'(\tilde r)}{4\tilde a(\tilde r)g(\tilde r)} + \frac{\tilde b'(\tilde r)g'(\tilde r)}{4\tilde b(\tilde r)g(\tilde r)} \nonumber \\
  &+& 
  \frac{\tilde f'(\tilde r)g'(\tilde r)}{4\tilde f(\tilde r)g(\tilde r)} + \frac{{g'(\tilde r)}^2}{2{g(\tilde r)}^2} +
\frac{g''(\tilde r)}{2g(\tilde r)} + 
  \frac{r^2\tilde A^{(0,1)}(\tilde r,z)\tilde C^{(0,1)}(\tilde r,z)}{L^2\tilde a(\tilde r)\tilde f(\tilde r)g(\tilde r)} + 
  \frac{3r^2{\tilde C^{(0,1)}(\tilde r,z)}^2}{L^2\tilde a(\tilde r)\tilde f(\tilde r)g(\tilde r)} \nonumber \\
  &+& 
  \frac{r^2\tilde C^{(0,2)}(\tilde r,z)}{L^2\tilde a(\tilde r)\tilde f(\tilde r)g(\tilde r)} +
\frac{g'(\tilde r)\tilde A^{(1,0)}(\tilde r,z)}{2g(\tilde r)} - 
  \frac{\tilde C^{(1,0)}(\tilde r,z)}{r} + \frac{\tilde a'(\tilde r)\tilde C^{(1,0)}(\tilde r,z)}{2\tilde a(\tilde r)} + 
  \frac{\tilde b'(\tilde r)\tilde C^{(1,0)}(\tilde r,z)}{2\tilde b(\tilde r)} + \frac{\tilde f'(\tilde r)\tilde C^{(1,0)}(\tilde r,z)}{2\tilde f(\tilde r)}  \nonumber \\
&+&
  \frac{7g'(\tilde r)\tilde C^{(1,0)}(\tilde r,z)}{2g(\tilde r)} + \tilde A^{(1,0)}(\tilde r,z)\tilde C^{(1,0)}(\tilde r,z) + 
  3{\tilde C^{(1,0)}(\tilde r,z)}^2 + \tilde C^{(2,0)}(\tilde r,z)            =0.
\label{eqnu3}
\end{eqnarray}
The index $(n,m)$ above the functions $\tilde A,\tilde B,\tilde C$ refer to the order of the derivative in $\tilde r$ and $z$ respectively while the primes over the functions $\tilde f,\tilde a,\tilde b$  denotes the derivative with respect to $\tilde r$. The function $g(\tilde r)$ is defined as $g(\tilde r)=r_h^2 + \tilde r^2$.

We integrated the system of non-linear partial differential equations for some values of the cosmological constant with the solver Fidisol \cite{fidi} based on a Newton-Raphson method, supplemented with the following boundary conditions 
\bea
\partial_{\tilde r}\tilde A(0,z)=0\ ,\ \partial_{\tilde r}\tilde C(0,z)=0, \tilde B(0,z) - \tilde A(0,z) = d_0 \nonumber\\
\tilde A(\infty,z) = 0\ ,\ \tilde B(\infty,z) = 0\ ,\ \tilde C(\infty,z) = 0,\\
\partial_z \tilde A(\tilde r,0)=0\ ,\ \partial_z \tilde B(\tilde r,0)=0\ ,\ \partial_z \tilde C(\tilde r,0)=0,\nonumber\\
\partial_z \tilde A(\tilde r,L)=0\ ,\ \partial_z \tilde B(\tilde r,L)=0\ ,\ \partial_z \tilde C(\tilde r,L)=0\nonumber,
\label{bcnubs}
\eea
$L$ being the critical length given by the first order analysis, $L=2\pi/k$. These boundary conditions are similar to the one used in the asymptotically locally flat non uniform black string problem \cite{rkk,sorkinnu}.

The parameter $d_0 $ is related to the temperature of the non uniform black hole $T^{NU}_H$ according to 
\be
T_H^{NU} = e^{-d_0 } T_H^U,
\ee
$T_H^U$ being the temperature of the background uniform solution. 

Note that here we keep the length $L$ fixed and vary the temperature, leading to a non uniform solution with the critical length and different temperature than the background uniform solution. Note also that the entropy $S^{NU}$ changes along the non uniform branch and is given by
\be
S^{NU} = S^U \int_0^L e^{\tilde B(0,z)+\tilde C(0,z)}dz,
\ee 
$S^U$ being the entropy of the background uniform phase.

\begin{figure}
	\centering
		\includegraphics[scale=.5]{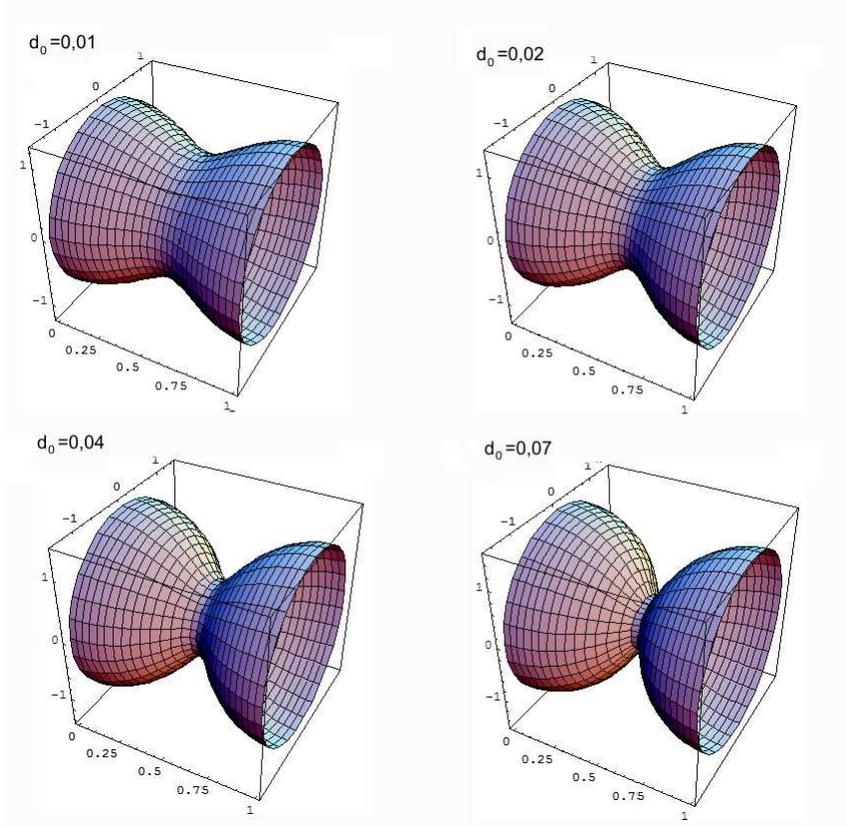}
		\caption{Embedding of the horizon surface in a three dimensional Euclidean space for $r_h=1,\ \Lambda=-0,1$. Each points on the surface represents a $S_3$ sphere. The equation of the surface in cylindrical coordinate $(\rho,\theta,z)$ is given by $\rho = r_h e^{C(r_h,z/L)}$.}
	\label{fig:immersed}
\end{figure}

By construction, equations \eqref{eqnu1}, \eqref{eqnu2} and \eqref{eqnu3} admit the trivial solution $\tilde A = \tilde B = \tilde C = 0$ for $d_0  = 0$. This was checked numerically providing a crosscheck of our equations. We started with the trivial solution as an initial guess and gradually increased the value of $d_0 $, but the solver failed to provide a convincing solution. Instead, starting with a combination of the trivial solution and of the first order solution and a small but non vanishing value of the parameter $d_0 $ leads to a consistent results, which are numerically robust. It has been checked that in the limit where $d_0 $ goes to zero, the trivial solution is recovered, connecting the non-uniform black string phase to the uniform black string. 
When $d_0 $ is increased, the solutions develop a more and more pronounced extremum for $z=L/2$, suggesting a transition to a localised black hole; this is shown in figure \ref{fig:immersed} where we plot an embedding of the horizon in $\mathbb R^3$ for $\Lambda=-0,1$ and different values of $d_0 $.

\begin{figure}
	\centering
		\includegraphics[scale=.6]{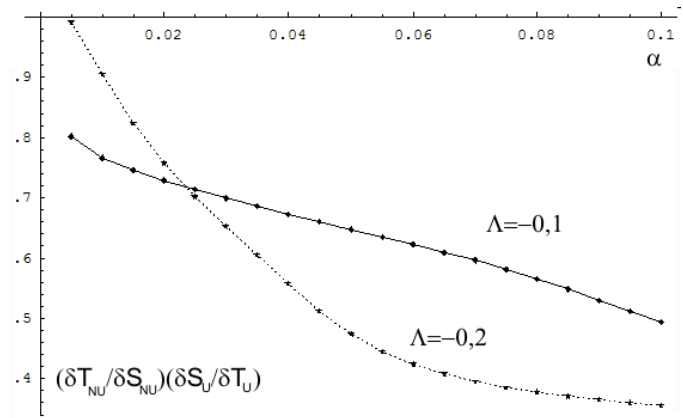}
	\caption{The ratio of the specific heat ($\delta S/\delta T_H)$ computed with the perturbative approach and with the non perturbative approach. In the figure, $NU$ refers to the non perturbative solution while $U$ refers to the perturbative solution. For small values of $d_0 $, the ration is close to one.}
	\label{fig:dtdscomp}
\end{figure}

We compared the value of the corrections on the horizon quantities coming from the perturbative approach and the same values computed from the non perturbative approach and it appeared that our results are consistent, providing a crosscheck of both the perturbative and non perturbative solutions. This is shown in figure \ref{fig:dtdscomp} for three values of the cosmological constant. 

\begin{figure}
	\centering
		\includegraphics[scale=.5]{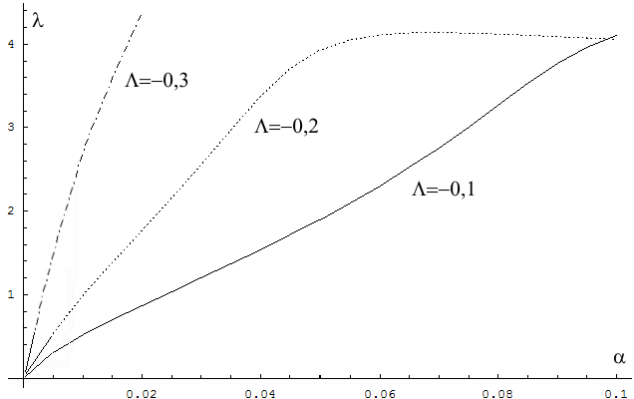}
	\caption{The value of $\lambda$ as a function of the parameter $d_0 $ for $r_h=1$, $L=2\pi/k$, where $k$ is the critical wavenumber and different values of $\Lambda$. The value of $\lambda$ increases quicker with $d_0 $ for larger negative values of the cosmological constant.}
	\label{fig:def}
\end{figure}

We computed the deformation parameter \cite{gubser} $\lambda=R_{max}/R_{min} - 1$, with $R_{max} = max(r_h e^{C(r_h,z)})$, $R_{min} = min(r_h e^{C(r_h,z)})$. It appeared that the value of $\lambda$ increases quicker with $d_0 $ for larger negative values of the cosmological constant; this is shown in figure \ref{fig:def}.

It must be stressed that the numerical investigation was plagued by several difficulties, in particular, the solver was very sensitive to the mesh and failed to provide reasonable solutions once the value of the cosmological constant is large, this can be due to the fact that the value of the background solution becomes very high for large values of the cosmological constant, forcing the program to deal with extremely small numbers for the non uniform fields (recall that they go to zero asymptotically) and extremely large numbers for the background fields (which diverge like $r^2$). This is the reason why we could even with many difficulties investigate the solution up to only $\Lambda=-0,3$. Note also that the precision of the solutions is of the order of $5\%$ with a $40\times 81$ mesh for $z\times \tilde r$, uniform in the $z$ direction; it is possible to reduce this error by considering a denser mesh. However, it has been checked that the results presented here are compatible with results obtained from a denser mesh (within the numerical error). Let us finally mention the fact that the mass and tension can be in principle computed since the asymptotic of the full solution should follow the Fefferman Graham expansion. However, it turned out that the useful subleading coefficients were impossible to extract in our case. However, it should be possible to integrate the Smarr relation and the first law, but we didn't have data enough in order to extract precise values of the mass and tension.

\section{Conclusion}
We have integrated the non uniform black strings equations in $AdS$ up to second order in perturbation theory for many number of dimensions. We investigated the thermodynamical properties of these perturbative non uniform solutions and found evidences for the existence of a new stable phase of non uniform black strings. This new phase has the property that the non uniform black string length is of the order of the $AdS$ radius. We propose the terminology long and short $AdS$ black strings in contrast with the small and big $AdS$ black strings terminology. The small (resp. large) black strings are characterised by a small horizon radius - $AdS$ length ratio (resp. of order one and larger); the short (resp long) non uniform black strings are characterised by a small length in the extra direction - $AdS$ radius ratio (resp. of order one and larger). This new phase is not present in the uniform case where all the thermodynamical quantities can be defined per unit length, so that the length plays a spectator role. For non uniform black strings, the length of the extra direction appears in a non trivial way in the definition of the thermodynamical quantities.

We were not able to investigate the problem of ($\ell$-depending) critical dimension here (see \cite{sorkin} for asymptotically locally flat spacetime). The problem comes from the extraction of numerical factors needed in order to construct the conserved global quantities. A possible alternative was to construct these quantities by integrating the first law of thermodynamics; this works qualitatively well except in the region of small horizon radius-$AdS$ length ratio. Unfortunately, this is precisely the region needed. A reconsideration of the numerical technique is needed in order to solve this problem. 

However, we believe these new phases of non uniform black string to be strongly related to the occurrence of critical dimensions in canonical ensemble. Moreover, the change of sign in the temperature correction should translate in a change of sign in rescaled mass correction in a rescaled mass-relative tension diagram; resulting if true in a connection with critical dimension in microcanonical ensemble.

Another interesting aspect of these new phases is the connection with the boundary conformal field theory. Our results suggests the existence of a stable non uniform long configuration in a $S^1\times S^{d-3}$ spatial background in the CFT side.

We also gave some preliminary results on the full nonlinear problem's solution. Despite of our efforts, we faced many difficulties in the investigation of the solutions, especially for larger negative values of the cosmological constant. However, our non perturbative results confirm the perturbative solutions, and indicate a possible transition to a localised black hole system. Unfortunately, we could not check the existence of the thermodynamically stable phase, predicted by the perturbative analysis. Clearly, more efforts should be done and a systematic investigation of the set of parameter should be investigated in the non perturbative regime.

Finally, let us give some interesting perspectives to this work. One is the hunt of critical dimensions. An interesting question to address would be to investigate whether these sable long non uniform black strings could be the endpoint of the unstable $AdS$ black string with corresponding length decay. This should be numerically easier than problems involving topology changes and might be of peculiar relevance in the context of $AdS$/CFT correspondence.

\section{Acknowledgement}
I would like thank Yves Brihaye, Eugen Radu and Georges Kohnen for useful discussions and wise advices.



\end{document}